\documentstyle[proceedings,bibstyle]{crckapb} 

\newcommand{\half}      {\hbox{${1}\over{2}$}} 
\newcommand{\threehalf} {\hbox{${3}\over{2}$}} 
\newcommand{\fivehalf}  {\hbox{${5}\over{2}$}} 

\begin{opening}
\title{VERY COLD GAS AND DARK MATTER}

\author{F. COMBES}
\institute{DEMIRM, Observatoire de Paris\\
	61 Av. de l'Observatoire, F--75014 Paris, France}
\author{D. PFENNIGER}
\institute{Geneva Observatory, University of Geneva\\
	CH-1290 Sauverny, Switzerland }

\end{opening}

\runningtitle{VERY COLD GAS AND DARK MATTER}

\begin{document}

\begin{abstract}
We have recently proposed a new candidate for baryonic dark matter:
very cold molecular gas, in near-isothermal equilibrium with the
cosmic background radiation at $2.73\,\rm K$. The cold gas, of
quasi-primordial abundances, is condensed in a fractal structure,
resembling the hierarchical structure of the detected interstellar
medium. 

We present some perspectives of detecting this very cold gas, either
directly or indirectly. The H$_2$ molecule has an ``ultrafine"
structure, due to the interaction between the rotation-induced
magnetic moment and the nuclear spins. But the lines fall in the km
domain, and are very weak. The best opportunity might be the UV
absorption of H$_2$ in front of quasars. The unexpected cold dust
component, revealed by the COBE/FIRAS submillimetric results, could
also be due to this very cold H$_2$ gas, through collision-induced
radiation, or solid H$_2$ grains or snowflakes.  The $\gamma$-ray
distribution, much more radially extended than the supernovae at the
origin of cosmic rays acceleration, also points towards and extended
gas distribution.
\end{abstract}

\section{Introduction}
The possibility that most of the mass of the Universe could be under
the form of gas around or in between galaxies has been widely
discussed in the 1960's (e.g. the review by Peebles, 1971). At that
time gas was assumed to be distributed in a smooth homogeneous
fashion. This was not justified by observations of the interstellar
gas already then.  The intergalactic material was supposed to be
hydrogen exclusively (with some helium), either atomic, molecular,
ionized, or even in a condensed snow. Self-gravity was ignored.
Several tests were proposed, such as emission or absorption of HI at
the $21\,\rm cm$ line, the Gunn-Peterson test in the Ly$\alpha$ line
of HI or the Lyman band in molecular hydrogen; the existence of an
intergalactic plasma would have been detected through free-free
emission or absorption, recombination lines, chromatic phase lag, or
Thomson scattering.

All the tests were negative, constraining the density of any
intergalactic gas several orders of magnitude below the closure
density.  However, the homogeneity hypothesis is drastic, especially
when comparing to the now much better known ISM, and including gravity
as a major force.  The Gunn-Peterson test is of course invalidated in
case of a clumpy medium.

Even until recently, the hypothesis of cold gas as dark matter was
eliminated quickly, without critical reflection, through stability
arguments (e.g. Hegyi \& Olive 1986): for an homogeneous galactic
gaseous halo, the virial temperature is of the order of $10^6\,\rm K$;
the gas cannot remain hot, it would have been seen in X-ray; in any
case, at this temperature, cooling processes are violent, the gas
collapses and forms stars.  Hydrogen snowballs, massive enough not to
collide with each other ($m> 1\,\rm g$) quickly evaporate, unless
gravitationally bound (but then they join the problem of brown
dwarfs).

All the above objections disappear when a realistic hierarchical
structure of cold and dense clouds is considered, closely resembling
the familiar fractal structure of the detected ISM (Pfenniger et
al.~1994). A model was then built to account for the baryonic dark
matter around galaxies, composed of basic ``clumpuscules" of molecular
hydrogen, of a Jupiter mass (10$^{-3}$ M$_\odot$), but with a much
smaller density than brown dwarfs, with $10^{9-10}\,\rm cm^{-3}$ and
radius of $\approx 20\,\rm AU$ (Pfenniger \& Combes 1994). For these
basic units clumpuscule are self-gravitating, statistically in
equilibrium with pressure forces, at the limit of the adiabatic
regime, since then the radiation transfer time equals the dynamical
time.  They compose the smallest scale of a hierarchical structure,
that ranges over six orders of magnitude, up to giant molecular clouds
of $100\,\rm pc$; above this scale, bigger gaseous complexes are torn
apart by galactic shear. The ensemble of clumpuscules is in quasi
isothermal equilibrium with the bath of photons of the cosmic
background radiation at $T=2.73 {\,\rm K} (1+z)$.  Due to the fractal
structure (of dimension $D<2$), the clumpuscules collide together
frequently, with, for such fractal dimension, a rate of the order of
the cooling-heating and dynamical times.

But all the fractal structure is far from local thermodynamical
equilibrium as explained in another paper in this volume, the
equilibrium including gravity is only statistical.  The large
fluctuations prevent most of the clumpuscules to cool down and
collapse further (were they distributed homogeneously they would not).
Through these collision induced fluctuations, the gas maintains
exchanges with the background, coalescing in giving back energy, or
fragmenting and evaporating in absorbing energy.  The condensed
structure resembles closely the well-studied ISM gas, already known as
a $D<2$ fractal-like structure over several orders of magnitude in
scale (Larson 1981; Scalo 1985; Falgarone et al.~1992).  The main
difference is that star-formation in the visible disk has
metal-enriched the medium that can then cool down much faster, and the
heating sources have partly destroyed the condensed fractal and formed
a diffuse intercloud medium.

In galaxy outskirts, the condensed H$_2$ phase is almost only in
contact with the intergalactic radiation field, which photodissociates
a small fraction of it into HI gas, because at the envisaged column
densities ($\sim10^{25}\,\rm cm^{-2}$) H$_2$ can self-shield easily.
An even smaller fraction could be ionized.  Since the average surface
density of the gas is falling as $R^{-1}$, HI must disappear into
ionised gas at a critical radius; this corresponds to the sudden fall
off of HI measurements, resembling an ionisation front (e.g.~the case
of NGC 3198, van Gorkom et al.~1987, unpublished). From this point of
view, the atomic gas in the outer parts serves as a tracer of dark
matter.

There is indeed some evidence that the gas and dark matter are
intimately related.  From the flat rotation curves, the surface
density of dark matter $\sigma_{\rm DM}$ varies asymptotically as
$R^{-1}$, and as well the HI surface density $\sigma_{\rm HI}$
(cf.~Fig.~1). Bosma (1981) was the first to notice a constant ratio of
$\sigma_{\rm DM}/\sigma_{\rm HI}$ as a function of radius in spirals,
which has been confirmed by many authors (Sancisi \& van Albada 1987;
Puche et al.~1990), and varies between 10 and 20 according to the
morphological type (Broeils 1992; Carignan, this volume).

\begin{figure}
\vspace{5cm}  
\caption{Ratio of HI to Dark Matter (DM) surface densities in spiral
galaxies, adapted from Bosma (1981), Puche et al (1990) and Freeman (1992). }
\end{figure}

The presence of large gaseous extensions around galaxies can explain
the widespread detection of absorption lines in front of quasars. The
Ly$\alpha$ forest, and the large frequency of absorptions on a single
line of sight (up to 100) remain unexplained. If these absorptions
come from gas around galaxies, the derived cross-section of a galaxy
corresponds to a radius of $480\,\rm kpc$ at $z=2.5$ (Sargent
1988). The gas corresponding to these atomic absorptions remains a
small fraction of the critical density.  But already the total
contribution of the gas in damped Ly$\alpha$ systems amounts to about
the luminous matter density in present galaxies (e.g.~Lanzetta et
al.~1991).

The model of cold gas as dark matter sheds also some light on puzzles
such as the presence of huge amounts of hot gas in clusters (the hot
gas represents 8 to 10 times the mass in galaxies in some rich
clusters, Edge \& Stewart 1991), the overabundance of gas in
interacting galaxies (Braine \& Combes 1993), the necessity of gas
infall for maintaining star-formation and the spiral structure, and
the evolution of galaxies along the Hubble sequence (e.g.~Pfenniger et
al.~1994).

Recent models have also been proposed, taking up the hypothesis that
H$_2$ gas could contribute significantly to the dark matter, based
mainly in massive proto globular cluster clouds possibly mixed with
brown dwarfs (de Paolis et al.~1995; Gerhard \& Silk 1995).  In this
case, the cold H$_2$ is not {\it very\/} cold, but has a temperature
between 5 and $20\,\rm K$, which makes it easier to detect in
emission. 
To us is unclear, however, how cold gas coexisting with a brown dwarf
cluster could be maintained in the mutual gravity at this low
temperature, since the brown dwarfs must be subject to a significant
dynamical friction, and the cluster must undergo core collapses that
inevitably relax and virialize also the gas to a higher temperature.

\section{Local dark matter and gravitational stability }
\subsection {Fit of the rotation curve}
The rotation curve of our own Galaxy is well-known from several
tracers, including the HI, H$_2$ gas or ionised gas, and is typical
for a SBbc galaxy: massive bulge traced by a central peak of
rotational velocity, flat rotation in the outskirts, although
uncertainties become large outside of the solar circle, due to the
lack of precise distance indicators (e.g.~Fich \& Tremaine
1991). Given all the observed parameters of the bulge, stellar and
gaseous disks, is it possible to constrain the model of dark matter
confined in the disk, in particular in solar neighbourhood?

We have retained a simple axisymmetric model of the Galaxy, composed
with a bulge (Plummer component), an exponential stellar disk, with or
without truncation, and two gaseous disks.  The first represents the
observable molecular ring, and is modelled by a difference between two
Toomre disks of index $n=2$ (Toomre 1963). The second represents the
observed HI component, and is modelled by an empirical ring like
distribution, deficient in the center, and falling as $R^{-1}$ at
large radii. The mass of this component has been multiplied by an
adjustable ratio $\sigma_{\rm DM}/\sigma_{\rm HI}$ to model the
gaseous dark matter.  The disks are modelled with a finite thickness,
and the rotation curves are calculated following the integral
formulation from Casertano (1983).  The final fit is displayed in
Fig.~2, together with the observed points.  All the adopted parameters
are listed in Table 1.

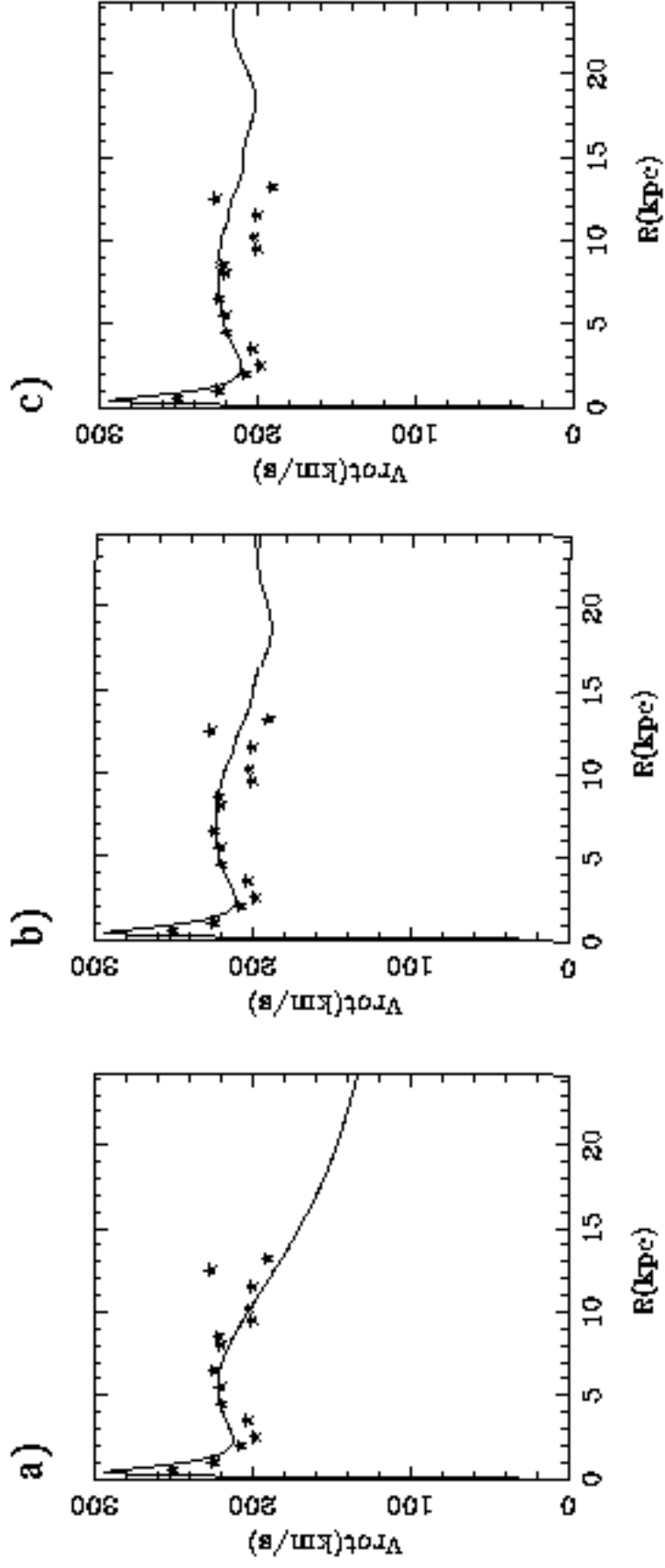
\begin{figure}
\vspace{5cm}  
\caption{Fit of the Milky Way rotation curve, with the parameters of
Table 1. Data points are from the compilation of Fich \& Tremaine
(1991).  The dark matter is assumed to follow the HI distribution: a)
without dark matter; b) with dark matter, with a distribution
following that of HI, and $\sigma_{\rm DM}/\sigma_{\rm HI}$ = 15; c)
same with $\sigma_{\rm DM}/\sigma_{\rm HI}$=20}
\end{figure}

\begin{table}[b]  
\vspace{-4mm}
\caption{ Parameters of the rotation curve fit }
\begin{center}
\begin{tabular}{lllll}
\hline
Component & Mass [$10^{10}\,\rm M_\odot$] & 
\multicolumn{2}{l}{Scale-lengths [kpc]} 
	& Type \\
\hline\noalign{\smallskip}
Bulge & $M_b = 1$ & $a_b=0.25$  &   & Plummer \\
Disk & $M_d = 6$ & $a_d = 2.7$  &   & Exponential \\
H$_2$ & $M_{\rm H_2} = 0.2$ & $R_1=6.0$  & $R_2=6.8$  
	& diff $n=2$ Toomre disks \\
HI & $M_{\rm HI}=0.4$ & $R_0 = 8.\ ^{(1)}$  &   & $\mu \propto R^{-1}$\\
\hline
\end{tabular}
\end{center}
$^{(1)}$ Start of the $R^{-1}$ behaviour
\end{table}

First, we see that it is possible to account for the rotation curve
without any dark matter until $R=10\,\rm kpc$.  Afterwards dark matter
is actually necessary to avoid a fall off of the circular speed
(Fig.~2a).  But adding the dark matter in the disk, with the bulk of
it outside the solar radius reduces the rotation speed inside
$R=10\,$kpc. The modelisation is therefore not unique, it depends
on several free parameters of total mass and concentrations, much more
than in the hypothesis of a spherical DM halo. For instance the
rotational velocity at the extreme galactic radii depends on how the 
HI-DM disk is truncated. Fig.~2b and c display reasonable good fits,
with  $\sigma_{\rm DM}/\sigma_{\rm HI}$=15 and 20 respectively.
We can estimate the dark matter
fraction in the disk at the solar radius at about 50\% of the total
mass.  This is compatible with the recent claim from Bahcall et al (1992),
but not from the Kuijken \& Gilmore studies (1989), but the question
of the existence of local dark matter is still open.

\subsection {HI scale-height}
The gaseous plane of the Milky Way is flaring linearly with radius,
outside the solar circle, i.e. the HI scale-height $h_z \approx 0.045
R$ (Merrifield 1992). If the gas is considered in gravitational
equilibrium in the direction perpendicular to the plane, with a
constant velocity dispersion $\sigma \approx$ 10 km/s as in face-on
galaxies (e.g. Dickey et al 1990), then different scale-heights are
predicted by the different models for the dark matter. In the case of
a spherical halo, the equilibrium requires $G M(R) R^{-3} h_z^2 \simeq
\sigma^2$, and since $M(R) \propto R$, we predict $h_z = 0.09 R$,
which is larger than observed.  A flattened dark halo is needed.  In
the case of a self-gravitating disk, where the dark matter follows the
HI flaring, the equilibrium requires $2\pi G \mu h_z \simeq \sigma^2$,
with the surface density $\mu \propto R^{-1}$; a height $h_z = 0.03 R$ is
predicted with the rotation curve model of the previous section. This
is quite compatible with the observations, given the uncertainty in
$\sigma$, and the probable overestimation of the observed height by the Milky
Way warp.

\subsection{Outer disk stability}
For Gerhard \& Silk (1995), the main problem raised by the existence
of a self-gravitating gaseous disk in the outskirts is its global
stability.  For a flat rotation curve, the surface density $\Sigma$
falls off as $R^{-1}$ as well as the epicyclic frequency $\kappa$, so
that the critical velocity for axisymmetric stability
$c_r\propto\Sigma/\kappa$ is constant.  This critical velocity
dispersion is of order $60-70\,\rm km\,s^{-1}$ in the Galaxy, while
the observed velocity dispersion {\it perpendicular\/} to the plane is
of the order of $7-10\,\rm km\,s^{-1}$ (e.g.~Dickey et al.~1990).

We have already discussed how this can be approached (e.g.~Pfenniger
et al.~1994).  First, gaseous disks are not razor thin: they flare
with radius (Merrifield 1992), which reduces self-gravity and the
critical velocity.  Second, real gaseous disks {\it are manifestly
unstable}, as witnessed by the spiral waves, asymmetries and large
scale inhomogeneities of outer HI disks in {\it every\/} spiral.
Third, the dispersion, averaged over the gravitational scales up to
$\sim1\,$kpc and over a few rotation periods, is indeed close to the
critical velocity.  Finally, the clumpy gaseous dark matter component
may still retain, as stars, a substantial dispersion anisotropy,
contrary to a classical smooth gas; so the velocity dispersion, well
measured perpendicular to the plane, is not necessarily a good
indicator of the effective horizontal dispersion (see also Elmegreen,
this volume).

\section{Perspectives of detection }

The above model is not only a plausible and conservative hypothesis on
the nature of dark matter, but it is a falsifiable one, since a series of
observations can be carried out to confirm or refute our propositions.
We review below possible observational tests and try to select the
most promising ones.

\subsection{ The hyperfine structure of ortho-H$_2$ }
The hydrogen molecule can be found in two species, para-H$_2$, in
which the nuclear spins of the two protons are anti-aligned, and the
resulting spin $I=0$, and ortho-H$_2$ for which the total nuclear
spin is $I=1$, with the spins of the two protons parallel. The
rotation quantum number $J$ is even for para-H$_2$ and odd for
ortho-H$_2$. In the para ground state $J=0$, there is no hyperfine
splitting, but for the ortho $J=1$, three levels can be identified,
corresponding to $F=0, 1$ and $2$. This splitting comes from the
interaction of the nuclear spin magnetic dipole, with the magnetic
field created from the motion of charges due to rotation. To this
interaction, must be added the spin-spin magnetic interaction for the
two nuclei, and the interaction of any nuclear electrical quadrupole
moment with the variation of the molecular electric field in the
vicinity of the nucleus (Kellog et al.~1939, 1940; Ramsey
1952). Magnetic dipole transitions are possible for $\Delta F =1$,
i.e.~there are two transitions, $F=2-1$, and $F=1-0$. 
The wavelength of these two transitions have been measured in the
laboratory at 0.55 and $5.5\,\rm km$, or more precisely at frequencies
of $546.390\,\rm kHz$ and $54.850\,\rm KHz$ respectively for $F=1-0$
and $F=2-1$.

In fact this structure could be called ultrafine structure, since
it is several orders of magnitude below hyperfine structure
(cf.~Field et al.~1966). Since the interaction involves two nuclear
momenta, the splitting is proportional to $\mu_n^2$, the nuclear magneton,
while the hyperfine structure involves the product of
$\mu_n$, with the Bohr magneton $\mu_o = e h/(4\pi m c)$, where
$m$ is the electron mass. The ultrafine to hyperfine structure ratio
is therefore $\mu_n/\mu_o= m/m_p$, where $m_p$ is the proton mass.

\subsection{ The ortho-para ratio }
\vspace{-1mm}
Only the ortho-H$_2$ is concerned by the ultrafine structure. Normal
molecular hydrogen gas contains a mixture of the two varieties, with
an ortho-to-para ratio of 3, when the temperature is high with respect
to the energy difference of the two fundamental states ($171\,\rm K$).
At lower temperatures, the ortho-to-para ratio must be lower, if the
thermodynamical equilibrium can be reached, until all the hydrogen is
in para state at $T=0$. However, due to the rarefied density of the
ISM, the ortho-to-para ratio is frozen to the H$_2$-formation value.
Considerable densities are required for the ortho-to-para conversion,
which occurs in solid H$_2$ for instance. 

But the fractal gas must be seen as a dynamical structure far out of
thermodynamical equilibrium, not only with large density contrasts,
but also large temperature contrasts.  The HI is then the warm
interface, and H$_2$ the coldest component, in a mass ratio of about 1
to 10.  By continuity, H$_2$ forms from HI and vice versa with a rate
given by the clump collision time-scale at the scale corresponding to
the virial temperature of the transition.  This is in any case
relatively short: for $3000\,\rm K$, $D\sim1.7$, we estimate a
duty-cycle of transformation HI to H$_2$ of the order of
$10^{6-7}\,\rm yr$. 

Now the key role in ortho to para conversion in interstellar clouds is
the proton exchange reaction (${\rm H}^+ + {\rm H}_2(j=1) \to {\rm
H}^+ + {\rm H}_2(j=0)$, cf.~Dalgarno et al.~1973; Gerlich 1990). This
reaction can transform the ortho in a time-scale $5\cdot 10^{13}
n({\rm H}_2)^{-1/2}\,\rm s$, if the H$^+$ ions in dense clouds are
essentially due to cosmic ray impacts, with the ionising flux $\xi =
10^{-17}\,\rm s^{-1}$ characteristic of the solar neighbourhood. The
corresponding time-scale for a clumpuscule near the sun is $10\,\rm
yr$, and the ortho fraction is negligible, but at large distances in
the Galaxy outskirts, where the cosmic-ray flux falls to zero, we can
expect a significant part of ortho-H$_2$ in the cold gas.

\subsection{ Detectability of the H$_2$ ultrafine lines }
On Earth, the ionosphere is reflecting the long radio wavelengths, 
which is useful for long distance communications.  The
ionospheric plasma is filtering all frequencies below the plasma
frequency $\omega = e (4\pi n/m)^{1/2} \approx 100\,\rm MHz$.  It is
therefore necessary to observe from space. Even from space, the long
wavelength radiations are somewhat hindered by interplanetary or
interstellar scintillations (e.g.~Cordes et al.~1986). 

\subsubsection { Interstellar plasma }
In the ISM the plasma frequency can be estimated by $\nu_p = 9
n_e^{1/2}\,\rm kHz$, where $n_e$ is the electron density.  Since the
latter is in average of the order of $10^{-3}\,\rm cm^{-3}$, the
plasma frequency $\nu_p \approx 250\,\rm Hz$. Radiation of frequencies
below that value does not propagate in the medium. More exactly, since
the ISM is far from homogeneous, low-frequency radiation propagates in
rarefied regions, and is reflected and absorbed by denser
condensations.  For kilometric wavelengths, there is no problem of
propagation, but the waves are scattered due to fluctuations in
electron density. The electric vector undergoes phase fluctuations,
since the index of refraction is $(1-\nu_p^2/\nu^2)^{1/2}$, where
$\nu$ is the radiation frequency.  If the ISM is modeled by a gaussian
spatial distribution of turbulent clumps of size $a$, the scattering
angle can be expressed by $
\theta_{\rm scat} \approx 
  10^8 (L/a)^{1/2} \langle \Delta n_e^2\rangle ^{1/2} /\nu^2 
\ \rm {radian},
$
where $L$ is the total path crossed by the radiation (e.g.~Lang
1980). At a typical distance of $L = 3\,\rm kpc$, and for the
frequencies considered ($\approx 200\,\rm kHz$), $\theta_{\rm scat}$
is of the order of $1^{\circ}$. This means that higher angular
resolution should be inaccessible below $0.1\,\rm MHz$.  The
scintillation problem is therefore severe, and hinders the 
resolution of point sources, but still the 
galactic disk can be mapped.

The interplanetary medium produces somewhat less scattering, and the 
total order of magnitude remains unchanged. 

\subsubsection { Intensity of the H$_2$ ultrafine lines }
The radiation has a dipole matrix element proportional to $\mu_n^2$;
the line intensity is therefore much weaker than for usual hyperfine
transitions (magnetic dipole in $\mu_o^2$). Since the spontaneous
emission coefficient $A$ is proportional to $\nu^3$, the life-time of
a hydrogen molecule in the upper ultrafine states is much larger than
a Hubble time: $A \approx 10^{-32}\,\rm sec^{-1}$. It is then likely
that the desexcitation is mostly collisional. Even at the $3\,\rm K$
temperature, the upper levels are populated in the statistical weights
ratio. A weak radiation is therefore expected, but the
velocity-integrated emission ($\int T_a \,dv$) is ten orders of
magnitude less than for the HI line, for the same column density of
hydrogen.  The prospects to detect the lines are scarce in the near
future, since it would need an instrument of about 6 orders of
magnitude increase in surface with respect to nowadays ground-base
telescopes!  A solution could be to dispose a grid of cables spaced by
$\lambda/4 \approx 125\,\rm m$ on a significant surface of the Moon,
e.g.~an area of $(300\,\rm km)^2$.  This requirement could be
released, however, if there exists strong coherent continuum sources
at km wavelengths. The H$_2$ ultrafine line could then be detected
much more easily in absorption, with presently planned km instruments.

\subsection { The HD and LiH Transitions and Detectability }
HD has a weak electric dipole moment; it has been measured in the
ground vibrational state from the intensity of the pure rotational
spectrum to be $5.85 \pm 0.17\cdot 10^{-4}$ Debye (Trefler \& Gush
1968).  The first rotational level is at $\approx 130\,\rm K$ above
the ground level, the corresponding wavelength is $112\,\mu$. This
line could be only observed in emission from heated regions, and given
the very low abundance ratio HD/H$_2 \approx 10^{-5}$ and weak dipole,
does not appear as a good tracer of the cold gas.

The LiH molecule has a much larger dipole moment, $\mu = 5.9$ Debye
(Lawrence et al.~1963), and the first rotational level is only at
$\approx 21\,\rm K$ above the ground level.  The corresponding
wavelength is $0.67\,\rm mm$ (Pearson \& Gordy 1969; Rothstein
1969). The line frequencies in the submillimeter and far-infrared
domain have been recently determined with high precision in the
laboratory (Plummer et al.~1984; Bellini et al.~1994), and the great
astrophysical interest of the LiH molecule has been emphasized
(e.g.~Puy et al.~1993). A tentative has even been carried out to
detect LiH at very high redshifts (de Bernardis et al.~1993).  This
line is unfortunately not accessible from the ground at $z=0$ due to
H$_2$O atmospheric absorption. This has to wait the launching of a
submillimeter satellite, in which case it is a good candidate. The
abundance of LiH/H$_2$ is at most $\approx 10^{-10}$, in the absence
of photodissociation, and the optical depth should reach 1 for a
column density of $10^{12}\,\rm cm^{-2}$, or $N($H$_2) = 10^{22}\,\rm
cm^{-2}$, in channels of $1\,\rm km\,s^{-1}$.

\subsection { The H$_2^+$ hyperfine transitions }
The abundance of the H$_2^+$ ion is predicted to be less than
$10^{-11}$ to $10^{-10}$ in chemical models (e.g.~Viala 1986). But the
H$_2^+$ ion possesses an hyperfine structure in its ground state,
unfortunately in the first rotational level $N=1$. The electron spin
is $\half$, and the nuclear spin $I=1$, which couple in $F_2 = I + S=
\half$ and $\threehalf$; then $F= F_2 + N = \half$, $\threehalf$ and
$\fivehalf$. Five transitions are therefore expected, of which the
strongest is $F$, $F_2= \fivehalf, \threehalf \to \threehalf,\half$,
at $1343\,\rm MHz$ (Sommerville 1965; Field et al.~1966). At the
interface between the cold molecular gas and the
interstellar/intergalactic radiation field, one can hope to encounter
a sufficient column density of H$_2^+$. The excitation to the $E_u =
110\,\rm K$ level is problematic however.

\subsection { C and O pollution of the quasi-primordial cold gas}
As soon as there exist some metal enrichment from stellar
nucleosynthesis, cold gas could be traced by CO molecules, provided a
sufficient column density can be shielded from photodissociation. The
abundance [O/H] decreases exponentially with radius in spiral
galaxies, with a gradient between 0.05 and 0.1 dex/kpc (e.g.~Pagel \&
Edmunds 1981), and the $N($H$_2$)/I(CO) conversion ratio is
consequently increasing exponentially with radius (Saka\-moto
1996). There could be even more dramatic effects such as a sharp
threshold in extinction (at 0.25 mag) before CO is detectable (Blitz
et al.~1990), due to photo-dissociation.  We can then estimate until
which radius the dense clouds are likely to contain CO molecules, if
we assume that the opacity gradient follows the metallicity
gradient. Assuming the proportionality relation $N($H)$\approx 2\cdot
10^{21}\,\rm A_V\ atoms\ cm^{-2}\, mag^{-1}$ between the gas column
density and opacity in the solar neighbourhood (Savage et al.~1977),
and a column density of $10^{25}\,\rm cm^{-2}$ for the densest
fragments, their opacity A$_{\rm V}$ falls to 0.25 at $R \approx
60\,\rm kpc$, but of course the CO disappears at larger scales before.

The lack of heating sources is another effect hindering the detection
of molecular tracers in emission, far from star-formation regions. It
is impossible to detect emission from a cloud at a temperature close
to the background temperature. Only absorption is possible, although
improbable for a surface filling factor of less than 1\%.  Absorption
is biased towards diffuse clouds (intercloud medium) with a large
filling factor and a low density (and therefore a low excitation
temperature).  This is beautifully demonstrated in the molecular
absorption survey of Lucas \& Liszt (1994) in our Galaxy. But this
diffuse medium is preferentially depleted in CO at low metallicity.

Galaxy clusters is a privileged environment where the cold gas might
be metal-enriched. Galaxy-galaxy interactions progressively heat the
cold gas coming from individual galaxies; the virialised hot medium
(10$^{6-7}$ K) experiences a strong mixing, and is enriched by the
galaxy ejecta, to the observed intra-cluster abundance of $\approx
0.3\,\rm Z_\odot$. In the cluster center, where the density of hot gas
is high enough, a cooling flow is started, and the gas temperature
runs away down to the background temperature again, in a fragmented
structure (Pfenniger \& Combes 1994).  The cluster medium is therefore
multiphase, with a dense phase completely screened from the X-ray flux
(Ferland et al.~1994).  This accounts for the apparent complete
disappearance of gas in cooling flows, and may explain the high
concentration of dark matter in clusters deduced from X-ray data and
gravitational arcs (Durret et al.~1994; Wu \& Hammer 1993).  Many
authors have tried to detect this gas in emission or absorption,
either in HI (e.g.~Dwarakanath et al.~1995) or in the CO molecule
(e.g. Braine \& Dupraz 1994).  Maybe the best evidence of the presence
of the cooling gas is the extended soft X-ray absorption (White et
al.~1991).  The gas has a high surface filling factor ($\approx 1$),
and a column density of the order of N$_{\rm H} \approx 10^{21}\,\rm
cm^{-2}$.  The total mass derived is of the order of $10^{11}\,\rm
M_\odot$ over a $100\,\rm kpc$ region.  This diffuse phase corresponds
to the interface between the very cold molecular gas and the hot
medium. Part of the interface is atomic, part is ionised (as observed
H$\alpha$ filaments suggest).  Although the HI is not detected in
emission with upper limits of the order of $10^9-10^{10}\,\rm
M_\odot$, it is sometimes detected in absorption, when there is a
strong continuum source in the central galaxy. The corresponding
column densities are $>10^{20}\,\rm cm^{-2}$.

If the molecular clouds are cold ($T\approx 3\,\rm K$) and condensed
(filling factor $< 1\%$), it is extremely difficult to detect them,
either in emission or in absorption, even at solar metallicity. The
best upper limits reported in the literature ($N($H$_2)< 10^{20}\,\rm
cm^{-2}$, average over $10\,\rm kpc$ wide regions, but assuming
$T\approx 20\,\rm K$ and solar metallicity, i.e.~the standard
$N($H$_2$)/I(CO) conversion ratio, are perfectly compatible with the
existence of a huge cold H$_2$ mass (the conversion factor tends to
infinity when the temperature tends to the background temperature).

\subsection { UV H$_2$ absorption in front of quasars }
For the bulk of the gas at $T = T_{\rm bg}$, only absorption could be
detected.  Absorption in the vibration-rotation part of the spectrum
(in infrared) is not the best method, since the transitions are
quadrupolar and very weak. An H$_2$ absorption in Orion has been
detected only recently (Lacy et al.~1994) and the apparent optical
depth is only about 1\%. This needs exceptionally strong continuum
sources, which are rare.  Electronic lines in the UV should be more
easy to see in absorption.

Molecular hydrogen has been found in absorption in front of the
PKS\-0528-250 quasar by Foltz et al.~(1988), at a redshift $z=2.8$
where there is already a damped Ly$\alpha$ system; the column density
is not very high, $10^{18}\,\rm cm^{-2}$, with an estimated width of
$5\,\rm km\,s^{-1}$ and a temperature of $100\,\rm K$. Recognizing
H$_2$ absorption is not easy in the Ly$\alpha$ forest, and many
tentatives have remained inconclusive. A careful cross-correlation
analysis of the spectrum is needed in order to extract the H$_2$ lines
from the confusion.  Already Levshakov \& Varshalovich (1985) had made
a tentative detection, towards PKS0528-250, with some 13 coincident
lines among the Lyman and Werner H$_2$ bands. High velocity resolution
to reduce confusion would be helpful in the future to detect more
systems. Again absorption is biased towards diffuse gas in galaxies,
but in less than $f=1\%$ of cases, a damped H$_2$ system should be
detected. A molecular clumpuscule on the line of sight of a quasar
should produce a very wide saturated absorption, since the line would
be in the square-root section of the curve of growth, and nearby lines
should overlap (most of the UV continuum could be absorbed). If the
clumpuscule is in our own galaxy, temporal variations are expected
over a few months interval. It might be the most promising way to
detect the cold H$_2$ gas in the outer parts of galaxies.

\subsection{ Submillimeter Continuum }
The far-infrared and submillimeter continuum spectrum from $100\,\mu$
to $2\,\rm mm$ has been derived from COBE/FIRAS observations by Reach
et al.~(1995).  They show that in addition to the predominant warm
dust emission, fitted by a temperature of $T\approx20\,\rm K$, there
is evidence for a very cold component ($T = 4-7\,\rm K$), ubiquitous
in the Galaxy, and somewhat spatially correlated with the warm
component (see also Mather, this volume). The opacity of the cold
component, if interpreted by the same dust model, is about 7 times
that of the warm component.  It could correspond to those dense clumps
of gas, shielded from the interstellar radiation field, that have been
polluted by dust and heavy elements.

Schaefer (1996, and this volume) proposes that the cold dust component
detected by COBE/FIRAS might be due in fact to molecular hydrogen
emission, as collision-induced dipole transitions: in small aggregates
at very high density (a fraction of Amagat$ = 4\cdot 10^{18}\,\rm
cm^{-3}$), the H$_2$ gas can emit a continuum radiation, corresponding
to free-bound or free-free transitions of weakly bound H$_2$ dimers,
containing a large fraction of ortho-H$_2$.  The para-para complexes
do not produce the radiation, by symmetry. Schaefer finds a good fit 
for the COBE spectra if the dense H$_2$ clouds follow the HI
distribution in the outer parts of the Galaxy.

The weak-dipole radiation due to H$_2$ collisional complexes is an
interesting possibility to detect the presence of cold molecular
hydrogen.  Only exceptionally dense regions could explain the signal
detected by COBE, since the emission is proportional to the square of
the density (Schaefer 1994). The required density then imposes the
temperature ($T>11\,\rm K$), to avoid the transition to solid
molecular hydrogen.  Already we had remarked that at the present
cosmic background temperature of $T_{\rm bg0}=2.726\pm 0.01\,\rm K$,
the average pressure in the H$_2$ clumpuscules was about 100 times the
pressure of saturated vapour, and that probably a fraction of the
molecular mass might be in solid form (Pfenniger \& Combes 1994).
Already traces of 
H$_2$ snow flakes can improve the coupling with the CBR, but the large
latent heat of $110\,\rm K$ per H$_2$ molecule and the lack of
nucleation sites prevent a large mass fraction to freeze out.  The
condition of dimerization is then largely satisfied in the conditions
of the clumpuscules ($T\approx 3\,\rm K$, $n\approx 10^{10}\,\rm
cm^{-3}$).  We expect continuum radiation to be emitted and absorbed
by the H$_2$ collisional complexes, through collision-induced dipole
moment.  The absorption coefficient peaks around $\lambda =0.5\,\rm
mm$. The optical depth of each clumpuscule is however quite low $\tau
\approx 10^{-9}$ which makes such signature hardly detectable.

\section{Gamma-ray distribution }
Gamma rays ($\gamma$) of high energy come mainly from the interaction
of cosmic rays (CR) with the nucleons of the ISM (e.g. Bloemen 1989).
Many attempts have been made in the recent years to derive the radial
distribution of CR's from observation of $\gamma$'s, assuming that the
gas distribution is well known, derived from HI and CO
measurements. Two main problems arose from these derivations: the CR
distribution obtained has a very much smoother and extended 
radial dependence (scale-length $16\,\rm kpc$) than the assumed CR
sources (supernovae and stellar distribution, scale-length $4\,\rm
kpc$); and the H$_2$ mass derived from CO emission and a constant
H$_2$/CO conversion ratio in the Galactic Center, appears too high by
a factor at least 3 with respect to the $\gamma$ rays detected there
(Osborne et al.~1987).

The fact that the $\gamma$-ray distribution is much more extended
radially than the CR sources has been interpreted in terms of CR
diffusion (Bloemen 1989).  However the amount of diffusion is not well
known. CR particles are closely following the magnetic field lines,
due to their small gyration radius, and the field intensity is a
function of gas volumic density, so that it is intrinsically hard to
disentangle the CR and gas distributions.  The possibility of
convection in the halo, that can redistribute the CR's in a flatter
radial profiles, has been debated. Large halos are excluded however,
from the study of primaries, secondaries, and radioactive secondaries
as a function of energy (e.g. Weber et al.~1992).  Also the thickness
of the radio synchrotron halo has been derived around $3.6\,\rm kpc$
(e.g. Beuermann et al.~1985), giving the scale-height of CR electrons.

We have tried to fit the observed EGRET $\gamma$-ray distribution
assuming different axisymmetric models for CR and total gas
distribution.  Some of the models are displayed in Fig.~3. Due to the
low CR density in the outer parts of the Galaxy, the existence of
large amounts of cold gas there is compatible with the data. A more
detailed model, releasing the axisymmetry hypothesis, and taking into
account the actual $l-b-v$ diagrams of the HI and CO emissions will be
reported in a future work.

\begin{figure}
\vspace{5cm}  
\caption{Fit of the $\gamma$-ray distribution; a) radial distribution of
cosmic rays adopted, including diffusion; b) $\gamma$-ray l-b map obtained
with the observed HI and H$_2$; c) same but the HI density multiplied by 20 }
\end{figure}

\section{Conclusions}
The main argument for the existence of baryonic dark matter comes from
the constraints of the Big Bang nucleosynthesis, compared with the
observed abundances of primordial elements. The most recent estimates
find that the baryon density relative to the critical closure density
must lie in the range $\Omega_B = 0.01-0.09$ (Smith et al.~1993). The
visible baryons account for a much lower density, around $\Omega =
0.002$ (Persic \& Salucci 1992).  But the galactic dark matter could
still be entirely baryonic (e.g.~Carr 1995).  The recent micro-lensing
experiments conducted towards the Magellanic Clouds have revealed that
Machos can account for about
20\% of these dark baryons (Aubourg et al.~1993; Alcock et al.~1995),
although the constraint is loose.

Molecular hydrogen is one of the least exotic candidate (Pfenniger et
al.~1994). Present observations are not incompatible with this
hypothesis.  The main difficulty to detect very cold gas in emission
is its temperature close to the one of the cosmic background.  
However, we must search for observational tests to falsify the
proposition.  The detection of the ``ultrafine" structure of the
ortho-H$_2$ molecules at km wavelengths raises considerable
difficulties for the near future, but could be a means to fix the
$N($H$_2$)/I(CO) conversion ratio in our Galaxy.  The LiH rotational
lines will be easily detectable by submillimeter satellites.

Absorption lines detection might be the best way if the gas is indeed
very cold. Since the surface filling factor of the molecular clumps is
low ($f<1\%$), large statistics are required, but the perspectives are
far from hopeless. H$_2$ absorption in the Lyman and Werner bands has
already been identified in a damped Ly$\alpha$ system, at $z=2.8$. For
a clumpuscule in our own galaxy falling just on the line of sight of a
quasar, we expect a strongly damped and transient absorption over a
few months.

Finally, it is not excluded that the cold dust component detected by
COBE/FIRAS is tracing the cold H$_2$ component, limited to galactic
radii where the cold gas is still mixed with some dust.  Gamma ray
data could also be interpreted with the help of radially extended gas
distributions.

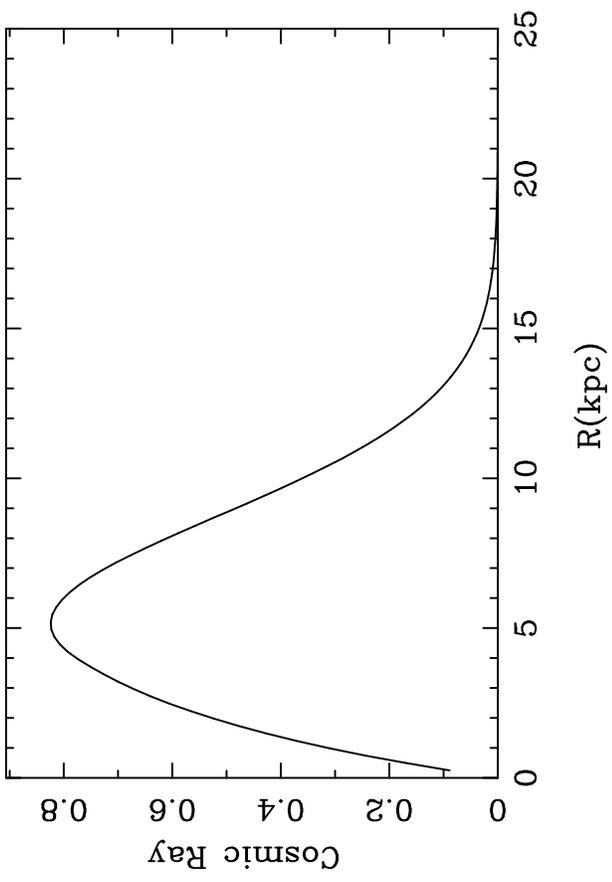
\end{document}